\documentstyle[preprint,aps]{revtex}
\begin{document}
\thispagestyle{empty}

\title{
 Low Energy Pion-Nucleus Interactions:  Nuclear Deep Inelastic 
Scattering, Drell-Yan and Missing Pions }

\author{Gerald A. Miller$^{a,b}$}

\address{Department of Physics$^a$
351560  and
Institute For Nuclear Theory$^b$
351550, University of Washington,
Seattle, Washington 98195, USA}

\maketitle

\begin{abstract}

The experimental discovery that the nucleus is approximately transparent to 
low energy pions is reviewed. The consequences of this for nuclear deep 
inelastic scattering and Drell-Yan interactions are discussed. I argue that 
low energy nucleus data imply that there is little nuclear enhancement of the 
pion cloud of a nucleon, and try to interpret this in terms of nucleon-nucleon 
correlations.

\end{abstract}
\vspace {20pt}
Text of invited talk presented at the LAMPF Users Group, Inc. Symposium, 
Oct. 25-26, 1996
\vspace{2.0cm}

\hfil {DOE/ER/41014-3-N97}\\

\hfil {DOE/ER/40561-299-INT96-19-06}\\

\newpage

\section{Introduction}

I will discuss low energy pion-nucleus interactions and the relationship 
between that subject and nuclear deep inelastic and Drell-Yan interactions and 
the problem of the missing pions.

I begin by mentioning my enthusiastic appreciation of two very successful 
parts of the Los Alamos Meson Physics Facility LAMPF program.  The first is 
the inelastic reactions in the $\Delta$ resonance region. The $\pi^+$ 
interact mainly with protons and the $\pi^-$ with neutrons. This feature was 
used to separate the proton and neutron contributions to inelastic transition 
densities at the nuclear surface. Many times the independently obtained proton 
contributions were consistent with those derived from electron scattering, 
thereby lending credence to the neutron results. One exciting spin-off was the 
ability to observe isospin mixing.

I also want to record my enthusiasm for the double charge exchange 
$(\pi^+,\pi^-)$ program. The experimentalists observed nucleon-nucleon 
correlations. These were not of the six quark bag type that I was excited 
about, but were correlations nonetheless!

Now I come to the outline of my own discourse, which concerns low energy 
($\approx$ 50 MeV) pion-nucleus interactions. LAMPF and TRIUMF
physicists discovered that 
the nucleus is transparent -- pions have essentially no multiple scattering in 
the nucleus. I discuss some of the evidence obtained from four different kinds
of measurments.
 It turned out that related 
studies of the $\pi$-nucleus interaction were being made at CERN in 
$(\mu,\mu')$ nuclear deep inelastic scattering. Nuclear effects on measured 
nucleon structure functions were observed -- the celebrated EMC effect. Models 
in which the nucleus enhanced the pion cloud of a nucleon, pion enhancement 
models (in which the pion bounced off many nucleons) were successful in 
reproducing the data. There were many other models which were able to 
reproduce the early data. Our response was to suggest that the nuclear 
Drell-Yan process pA $\rightarrow \mu^+\mu^- +X$ could be used to test such 
models - Los Alamos physicists responded by carrying out experiment 772 at 
FNLAB. One of the results is that there is no evidence for a pion excess. 
This is the famous missing pions problem. Perhaps this result should not have 
been surprising, in view of the low energy pion scattering measurements made 
at LAMPF and TRIUMF. I will then discuss a possible resolution of this problem.

\section{50 Mev $\pi^-$ Interaction}

The average $\pi$ nucleon cross section is about 25 mb at this energy. But the 
nuclear scattering requires sufficient momentum to push the bound nucleon 
above the Fermi sea. Including this feature leads [1] to an effective cross 
section of 15 mb, and a fairly long mean free path of 4 fm. Thus a pion 
travels a long distance as it moves through the nucleus.

The idea that low energy pions do not interact much is the community view. 
Here is a quotation from a well known text book [2]:

\begin{quote}
``In view of the relative weakness of the low energy interaction, the 
qualitative features of the $\pi$-nucleus scattering show up already in the 
first-order Born approximation."
\end{quote}

\noindent An example is the minimum caused by s-p interference in the elastic 
scattering angular distribution which has a position at an angle independent 
of nuclear target. This minimum is reproduced by using 
the first-order  Born approximation.

The relative weakness should have simplified various analyses. But the p-wave 
nature of the $\pi$N interaction caused big (erroneously computed) multiple 
scattering effects.

\section{P-wave Interaction} 

The P wave part of the average $\pi$N scattering amplitude 
$f(\vec{k}, \vec{k}')$ can be represented as
\begin{equation}
4\pi f (\vec{k}, \vec{k}') = \tilde{b}_1 \vec{k} \cdot \vec{k}'
\end{equation}
\noindent where $\vec{k}$ and $\vec{k}'$ are the final and 
initial relative 
pion-nucleon momenta. The optical potential $U$ is given the product of 
operators
\begin{equation}
U = 4 \pi f \rho \;\; ,
\end{equation}
\noindent which becomes the Kisslinger potential 
\begin{equation}
U(r) = b_1 \vec{\nabla} \cdot \rho (r) \vec{\nabla} \;\; ,
\end{equation}
\noindent when evaluated in coordinate space.

Consider the situation of the nuclear center where $\rho$ is approximately 
constant. Then the left-most gradient in Eq.~(3) can be pulled through to the 
right and the Klein Gordon equation reads
\begin{equation}
- \nabla^2 + b_1 \rho \nabla^2 \psi = k^2 \psi
\end{equation}
\noindent where $k$ is the on-shell wave number. One may find a solution of 
the form
\begin{equation}
\psi (\vec{r}) = e^{i \vec{K} \cdot \vec{r}}
\end{equation}
\noindent with
\begin{equation}
K^2 = {k^2 \over 1- b_1 \rho}\;.
\end{equation}

The above result does not look dramatic until one realizes that
\begin{equation}
b_1 \rho \approx 1
\end{equation}
\noindent so that $K^2$ is very large. This spurious effect can have bad 
calculational consequences. This is known as the Kisslinger singularity, but I 
want to stress that Leonard Kisslinger
did not encounter this problem in deriving his 
potential -- he worked wisely and correctly in the Born approximation.

Theorists were confronted by the problem of removing the singularity.

\section{Fixes of the Kisslinger Singularity}

Two different methods were used. The first was to modify the scattering 
amplitude or t-matrix for off-shell momenta:
\begin{equation}
4\pi f (\vec{k}, \vec{k}') = v(k) \vec{k} \cdot \vec{k}' 
v(k')
\end{equation}
\noindent with
\begin{equation}
\lim_{k \to \infty} v(k) = 0 \;\; ,
\end{equation}
\noindent as suggested by the finite size of the nucleon. The high momentum 
amplitude is cut off, and multiple scattering is suppressed as a result. There 
were many such models [3-6] of this kind, and the desire to obtain a 
quark-based $\pi$N interaction was a prime motivation for the cloudy bag 
model[7].

The second method invoked the repulsive nucleon-nucleon correlations--this was 
the EELL potential [8]. I will discuss this according to the argument 
presented in Ref.~9.

Consider the second order scattering operator $\hat{T}^{(2)}$ from two nucleons 
separated by $\vec{r}$ :
\begin{equation}
\hat{T}^{(2)} \sim\int \frac{d^3k}{(2\pi)^3} e^{i \vec{K}\cdot \vec{r}} 
\frac{[(\vec{k} \cdot \vec{K}\vec{K} \cdot \vec{k} - \frac{1}{3} k^2K^2) + 
\frac{1}{3} k^2K^2]}{k^2 - K^2 + i \epsilon}
\end{equation}
\noindent where virtual momentum K is the integration variable and k is the 
on-shell wave number. The factor $\vec{k}\cdot \vec{K} \vec{K} \cdot \vec{k}$ 
arises from two p-wave interaction vertices. A term $\frac{1}{3} k^2K^2$ has 
been subtracted from and added to the numerator to provide a decomposition 
into invariants. The two terms in the numerator can be shown to be very small. 
Integration over $d^3K$ leads to 
\begin{equation}
\frac{1}{3} k^2K^2 \rightarrow \frac{k^2}{3} [-\delta (\vec{r}) - k^2 
\frac{e^{ikr}}{4\pi r}] \;\; ,
\end{equation}
\noindent with the first term vanishing because of the correlations and the 
second of order $k^4$ and therefore small. The remaining term is a tensor. The 
integration over $\vec{K}$ gives
\begin{equation}
\vec{k}\cdot \vec{K} \vec{K} \cdot \vec{k} \rightarrow (\vec{k} \cdot \hat{r} 
k \cdot \hat{r} - \frac{1}{3} k^2) \;\;
\end{equation}
\noindent which after integration over $\vec{r}$ becomes
\begin{equation}
(\vec{k} \cdot \vec{\sigma}_1 \vec{k} \cdot \sigma_2 - \frac{1}{3} k^2 
\vec{\sigma}_1 \cdot \vec{\sigma}_2) \;\; ,
\end{equation}
which averages to zero. So there is very little second order or, by 
implication, higher order scattering.

Both methods lead to the result that there is little multiple scattering. It 
is interesting to recall that there were big fights over which method is 
better.  These days, everyone knows that Lagrangians can take on a variety of 
forms, generated by making field redefinitions. It is very likely that one can 
interpolate between the two methods with such a change of variables.

There is a physics question remaining. Although the multiple scattering is 
small--how small is it? Here is where the experimentalists came to the 
forefront by supplying beautifully relevant data.

\section{Single Charge Exchange on Proton and Nuclear Targets}

The reaction $\pi^- p \rightarrow \pi^0 n$ or $\pi^+ n \rightarrow \pi^0 p$ , 
equal by charge symmetry, was studied with the wonderful $\pi^0$ spectrometer.
Here
\begin{equation}
\frac{d\sigma}{d\Omega} (\pi^-p \rightarrow \pi^0n) = \frac{d \sigma}{d 
\Omega} (\pi^+n \rightarrow \pi^0p) = |A + B \vec{k} \cdot \vec{k}'|^2
\end{equation}
\noindent where A represents the repulsive Weinberg-Tomazowa S-wave term and B 
the attractive p-wave term. There is a tendency to cancel for forward
scattering angles with $\vec{k} \cdot \vec{k}'=k^2$ which becomes exact 
near 50 MeV. This cancellation is shown by the data displayed 
in Fig. 3 of Ref.10. The zero in the forward scattering amplitude
affords the opportunity to 
learn about multiple scattering in the initial $\pi^+$ and final $\pi^0$ state 
interactions. If such replaces $k^2$ by $K^2$ with $K^2 >> k^2$, the 
cancellation will not occur. The very same Fig. 3 of Ref.~10 
shows that the zero is 
maintained at approximately the same energy for nuclear targets throughout the 
periodic table. The simplest conclusion is that any multiple 
scattering must be very weak.

\section{Low Energy $(\pi_1 \pi')$ Interactions}

The reaction $\pi + A (O_1^+) \rightarrow \pi^+ + A^+ (O_2^+)$ also affords 
a study of multiple scattering. In Born approximation the scattering amplitude 
is the Fourier transform of the product of the initial and final wave 
functions. If the momentum transfer vanishes, orthogonality causes a vanishing 
scattering amplitude and a backward peaked cross section. (The momentum 
transfer is not exactly zero so that one does not expect an exact zero in 
the forward direction.)
Strong multiple scattering leads to a forward peak as seen 
in $(\alpha,\alpha')$ reactions. The $^{12}$C measurements [11,12], especially 
Fig. 2 of Ref. 11 indeed show  such a backward peaking, which is in 
agreement with an EELL calculation of Ref[11]. The angular distribution 
obtained from a Kisslinger potential which exactly reproduces the elastic 
data has a strong forward peak (see Fig. 11 of Ref.12) which disagrees with the 
inelastic 
data by about an  order of magnitude or more. The backward
 peaked angular distribution 
provides another signature for the approximate transparency of nuclei to 
pions.

\section{$\pi$-Nucleus Total Cross Sections}

Measurements of total cross sections provide a test of transparency or lack of 
multiple scattering. If only single scattering occurs, the total cross section 
will be proportional to the number of nucleons, A, in a nucleus. The 
experiments of Saunders, et al [13] indeed found such a relation:
\begin{equation}
\frac{\sigma_{tot} (\pi A)}{A} \approx \sigma(\pi N) \;\; .
\end{equation}

The results are plotted in Fig. 8 of Ref.13. The $\pi^+$ data satisfy 
Eq.~(15)
very precisely. The $\pi^-$ data show a slight rise with A which is presumably 
due to higher order effects of the attractive $\pi^-$- nuclear Coulomb 
interaction.

I have discussed evidence from elastic angular distributions, single 
charge exchange reactions, inelastic scattering and total cross sections. 
These indicate that the nucleus is approximately transparent to pions. This 
was discovered during the l970's, l980's and l990's.  It is interesting to 
realize that other investigations of the nuclear pion content were being 
carried out at CERN.

\section{EMC Effect}

The European Muon Collaboration (EMC) was involved with deep inelastic $\mu$ 
scattering experiments to determine the $F_2 (x)$ structure function of 
nucleons. Experiments were done on Fe nuclei to increase the cross section 
with the expectation that the nucleus looks like A free nucleons in these high 
$Q^2 \sim 100$ GeV$^2$ reactions. This was a common popular assumption. The 
data for the ratio of Fe to D structure functions showed otherwise. The 
structure function of a bound nucleon depends on its nuclear surroundings. The 
first data [14] showed that the structure function ratio was about 1.15 at x = 
0.05 and dropped approximately linearly with x until x reached $\simeq$ 0.65. 
See Fig. 2 of Ref.~14 which shows the significant systematic errors. This 
EMC effect was astounding. Nuclei affect quark distributions. Quarks were once 
and forever part of nuclear physics.

\section{Ericson-Thomas[15] Pion Enhancement Model}

It was natural for nuclear theorists to attempt to explain the EMC effect 
using nuclear mechanisms. I think the best such attempt was the work of 
Ericson and Thomas. Pions contribute to deep inelastic scattering from a 
nucleon because a nucleon can emit a virtual pion. The incident $\gamma^*$ 
then can bash this pion into bits.  This process can be enhanced in a nucleus 
because the virtual pion can multiple 
scatter amongst all the target nucleons. See Fig.1 of Ref.15. Intermediate 
nucleon and $\Delta$-hole states are formed. The repulsive particle-hole 
interaction is mocked up by the Landau-Migdal parameter $g'$. The virtual 
pions of momentum 300-400 MeV/c were strongly enhanced in that calculation, 
hence the name pion enhancement model. The resulting calculations, shown 
in Fig.~2 of Ref.~15 reproduced 
the early EMC data. The data of Refs.~10-13 were net yet available.

\section{EMC Means Everyone's Model is Cool}

There were many models that could reproduce the EMC effect. Examples are shown 
in Fig.1 of Ref.16. One could use six quark bags with structure functions 
different from those of free nucleons, pion enhancement models, linear 
combinations thereof, and dynamical rescaling which invokes the assumption 
that the scale parameters which governs perturbative QCD evaluation depends on 
the nucleus. All such provided a qualitative reproduction of the growing set 
of nuclear DIS data, except in the shadowing region which occurs at low x.

So there were many cool models, and it became necessary to derive tests to see 
which of these (if any) were really correct. My response was to suggest that 
Drell Yan $\mu^+\mu^-$ pairs be measured in high energy proton-nucleus 
interactions.

\section{Drell Yan Disentangles the EMC Effect [16]}

I was serving on a committee to develop a proposal for LAMPFII with a 
homework assignment of dreaming up new experiments. I thought that the Drell 
Yan process
\begin{equation}
p + A \rightarrow \mu^+\mu^- + X
\end{equation}
\noindent which occurs, for example, when a quark from the proton beam 
annihilates with an antiquark from the nuclear target to form a virtual time 
like virtual photon. The signature is the $\gamma^*$ 
decay into easily detectable $\mu^+\mu^-$ pairs.

Our calculations [16] showed that the different models, behaving similarly in 
deep inelastic scattering, acted very differently with the Drell Yan probe. 
There was an existing set up for Expt 605 already in place at Fermilab. The 
Los Alamos physicists were thus able to do Drell Yan measurements [17].

\section{Nuclear Dependence of Drell Yan Expt 772 [17]}

This experiment measured cross section ratios: C/$^2$H, Ca/$^2$H, Fe/$^2$H, 
W/$^2$H the result that there was no excess of pions or anything else. There 
was very little A dependence. See Fig. 3 of Ref. 17. The ratio for Fe/$^2$H was 
the one most studied theoretically and pion excess and quark cluster model 
calculations yielded ratios much larger than unity, in serious disagreement 
with the E772 data.  The paper, Ref.17, displayed an incorrect pion excess 
calculation, but an improved one, Fig. 3 
(with g'=0.7) of Ref.18, obtained a ratio of 1.1, also in 
disagreement with the data. The dynamical rescaling model appeared to agree 
with the Drell Yan data, but when used with the same parameters, disagrees 
with the deep inelastic scattering data [19].

No pion excess was seen. A similar result was obtained with the (p,n) and 
$(\vec{p},\vec{n})$ reactions. See Rappaport's paper at this conference.

This failure to observe the pions was a quite a puzzle. Bertsch, Frankfurt 
and Strikman wrote a paper ``Where are the nuclear pions," See Ref.~20.

\section{Why Pions Are Not Enhanced in Nuclei}

I shall present two arguments. The first is that we should have expected no 
enhancement, after 1985 or so, because 
the low energy pion-nucleus data showed that not 
much multiple scattering occurs. It is this very same multiple scattering 
which provided the pion enhancement in Ref.15. 
It is true that the pions in deep inelastic scattering are virtual 
and space like. But this should diminish the interactions since these pions 
are further away from the nucleon pole and $\Delta$ resonance. Thus we now
know that the low energy pion-nucleus
data imply that there should be little, if any, pion enhancement.

It would also be satisfying to understand this using the theory. Mark Strikman 
(PSU) and I are currently working on this problem. We wanted to start with some 
simple, understandable calculation.  We started by 
computing the pion excess per 
nucleon $\delta n_\pi$ in the deuteron. There are two contributions. The first 
is due to the change in the single nucleon term due to the binding. The energy 
denominator of the pion-two-nucleon intermediate state is larger than for the 
corresponding $\pi$-nucleon state, because of the extra kinetic energy of the 
spectator nucleon. The resulting contribution to the $\delta n_\pi$ is small 
and negative. This is a new term.

There is another term due to the pion exchanged between two nucleons; this 
gives a small positive contribution which is cancelled approximately by the 
new term mentioned above. The exchange term is small because the deuteron wave 
function does not support the exchange of a pion of momentum of about 400 MeV/c.

So there is almost no pion excess in the deuteron in our evaluation. One might 
argue that this would not occur for heavy nuclei.  Indeed 
there is no N$\Delta$ intermediate state in the deuteron, while such are 
allowed in heavy nuclei. However, the virtual $\Delta$ can not propagate a 
great distance and correlations should suppress their effects.  
Furthermore,  the short range 
correlations in nuclei are very similar to those of the deuteron wave 
function, and we expect to get a small pion excess in heavy nuclei.

We further speculate that any model calculation producing a large pion excess 
will also produce low energy pion nucleus calculations in disagreement with 
the low energy data.

I would like to discuss the fundamental difference between our deuteron 
multiple scattering 
calculations and those of Ref.15. One can compare diagrams of the same order 
to get an idea. In Ref.15 the repulsion is represented by a constant $g'$. 
The deuteron matrix elements contain 
factors of $\int d^3 p \psi^\dagger (\vec{p} - \vec{q}/ 
2) \cdots \psi(\vec{p} - \vec{q} / 2 ) \propto 
F(q)$. When two nucleons exchange a 
pion of momentum $\vec{q}$, this nuclear form factor cuts off the 
contributions for high $\vec{q}$. My belief is that no single parameter $g'$, 
no matter how well motivated, can do the job of a whole function. Indeed, this 
is one of the main mechanisms in another approach [20] towards understanding 
why the pions are not enhanced.

\section{Summary}

I list the major points.

$\bullet$ The $(\pi, \pi')$ and $(\pi^+, \pi^+, \pi^-)$ programs were major 
successes

$\bullet$ Low energy $\pi$ reactions show that the nucleus is transparent to 
pions

$\bullet$ This transparency is consistent with nuclear Drell Yan 
data

$\bullet$ Early $\pi$ enhancement predictions were based on good ideas, but an 
oversimplified theory.

There is still a problem. To my knowledge there is no theoretically viable 
model which explains both the nuclear deep inelastic and Drell Yan data.

I thank the
national Institute for Nuclear Theory for its support. 
This work is supported in part by the USDOE 
under Grants DE-FG03-97ER41014 and DE-FG06-88ER40427.

\end{document}